\documentclass[3p,times,authoryear]{elsarticle} %

\usepackage{ecrc}

\volume{00}

\firstpage{1}

\journalname{Journal Name}

\runauth{H. Meghnoudj, B. Robu, M. Alamir}

\jid{procs}

\jnltitlelogo{Clinical Neurophysiology}

\usepackage{amssymb}

\usepackage{xcolor}
\definecolor{lightblue}{RGB}{0,67,147}
\usepackage{lipsum}
\usepackage[colorlinks=true,			
			linkcolor=lightblue,
			citecolor=lightblue,
			urlcolor=lightblue]{hyperref}
\usepackage{marginnote}
\usepackage{amsmath}
\usepackage{relsize}
\usepackage{amsmath}
\usepackage{bm}

\makeatletter
\newcommand{\customlabel}[2]{%
   \protected@write \@auxout {}{\string \newlabel {#1}{{#2}{\thepage}{#2}{#1}{}} }%
   \hypertarget{#1}{}
}
\makeatother

 \DeclareMathOperator*{\argmin}{argmin}
\usepackage{mathtools}
 \DeclarePairedDelimiterX{\norm}[1]{\lVert}{\rVert}{#1}
\graphicspath{{./figures/}}
\usepackage{scalerel,stackengine}
\newcommand\pig[1]{\scalerel*[3.8pt]{\big#1}{%
  \ensurestackMath{\addstackgap[1.1pt]{\big#1}}}}

\usepackage{graphicx}
\newcommand{\smallU}{
  \mathchoice
    {{\scriptstyle\mathcal{U}}}
    {{\scriptstyle\mathcal{U}}}
    {{\scriptscriptstyle\mathcal{U}}}
    {\scalebox{.7}{$\scriptscriptstyle\mathcal{U}$}}
  }
\usepackage[ruled, vlined]{algorithm2e}
\usepackage{booktabs}
\usepackage{multirow}
\usepackage{float}
\usepackage{wrapfig}
\usepackage{enumitem}

\usepackage{titlesec}
\titlespacing{\subsection}{0pt}{*1}{*1}

\usepackage[figuresright]{rotating}

\begin{document}

\begin{frontmatter}



\dochead{}

\title{Sparse Dynamical Features generation, application to Parkinson's Disease diagnosis}

\author[label1]{Houssem Meghnoudj\corref{cor1}}
\address[label1]{Univ. Grenoble Alpes, CNRS, Grenoble INP, GIPSA-lab, 38000 Grenoble, France}
\cortext[cor1]{Corresponding author at: 11 Rue des Mathématiques, 38400 Saint-Martin-d'Hères, France\vspace{1.5pt}}
\ead{houssem.meghnoudj@gipsa-lab.fr}

\author[label1]{Bogdan Robu}
\ead{bogdan.robu@univ-grenoble-alpes.fr}

\author[label1]{Mazen Alamir}
\ead{mazen.alamir@gipsa-lab.inpg.fr}

\begin{abstract}
In this study we focus on the diagnosis of Parkinson's Disease (PD) based on electroencephalogram (EEG) signals. We propose a new approach inspired by the functioning of the brain that uses the dynamics, frequency and temporal content of EEGs to extract new demarcating features of the disease. The method was evaluated on a publicly available dataset containing EEG signals recorded during a 3-oddball auditory task involving \mbox{N = 50} subjects, of whom 25 suffer from PD. By extracting two features, and separating them with a straight line using a Linear Discriminant Analysis (LDA) classifier, we can separate the healthy from the unhealthy subjects with an accuracy of 90\,\% $(p < 0.03)$ using a single channel. By aggregating the information from three channels and making them vote, we obtain an accuracy of 94\,\%, a sensitivity of 96\,\% and a specificity  of 92\,\%. The evaluation was carried out using a nested Leave-One-Out cross-validation procedure, thus preventing data leakage problems and giving a less biased evaluation. Several tests were carried out to assess the validity and robustness of our approach, including the test where we use only half the available data for training. Under this constraint, the model achieves an accuracy of 83.8\,\%.
\end{abstract}

\vspace{-10pt}
\end{frontmatter}

\section*{Highlights}
\begin{itemize}[noitemsep,topsep=4pt,parsep=3pt]
\item Features derived from the dynamical, frequency and temporal content of EEGs are relevant biomarkers for the diagnosis of PD.  
\item Two explainable features are sufficient for an LDA model to achieve a classification accuracy of $94\,\%$.  
\item Few features with simple classifiers are more suitable for practical use, more explainable and trustworthy.   
\end{itemize}

\section{Introduction} \label{introduction}
\noindent Parkinson's Disease (PD) is a chronic neurodegenerative disorder affecting more than $6$ million persons worldwide as reported by the World Health Organization \citep{Who2016, Dorsey2018}. It is primarily caused by the lack of dopamine in the brain, due to the slow death of the dopaminergic cells \citep{Who2016, Balestrino2020}. PD is known by the general public for its motor symptoms such as: tremor at rest, rigidity, bradykinesia, akinesia, etc. \citep{Who2016, Balestrino2020}, however, non-motor symptoms may accompany or precede the onset of motor symptoms, sometimes even arriving $20$ years before the onset of the latter \citep{Chaudhuri2005, Kalia2015}. Affecting patients on a daily basis, non-motor symptoms are various: pain, fatigue, sleep disturbances, bradyphrenia, communication issues, etc. , only to cite a few \citep{Pfeiffer2016, Witjas2002}.\\

\noindent The diagnosis of PD is entirely clinical and is usually based on the manifestation of the motor symptoms \citep{Berardelli2013}. Before the appearance of the latter, patients suffer, raising the urge to look for new biomarkers allowing the early diagnosis of PD. The clinical diagnosis of PD is generally based on a pathological diagnosis or on a clinical diagnosis criterion (as the one from the United Kingdom PD Society Brain Research Center \citep{GibbWRG1998}). The overall clinical diagnosis accuracy is in the order of \mbox{$75$ \%} according to the World Health Organization \citep{Who2016} or around \mbox{$79$ \%} following \citep{Rizzo2016}. It is very important to note that the clinical diagnosis accuracy did not significantly improve during the last years particularly in the early stages of the disease where the response to dopaminergic treatment is not clear and less prominent \citep{Rizzo2016}. Actually, during the early disease manifestation ($<\!5$ years of disease duration) the clinical diagnosis accuracy is around \mbox{$53$ \%} and even lower, around \mbox{$26$ \%} accuracy, for patients with $<\!3$ years disease duration \citep{Adler2014}.\\
\noindent Electroencephalography (EEG) is a non-invasive method to record the electrical activity on the scalp which has been shown to represent the macroscopic activity of the brain underneath. It is used by several studies to assess individuals health conditions and to study brain function in healthy individuals as well as to diagnose various diseases that~alter the brain electrical activity such as: Parkinson's Disease, epilepsy, Alzheimer's, sleep disorders, schizophrenia, etc~\citep{Soufineyestani2020}.\\

\noindent EEG signals are known to have a low signal-to-noise ratio and present many difficulties. EEG noise is defined by any measured signal whose source is not the coveted brain activity \citep{Uriguen2015}. Unfortunately, in most cases the EEG signal is contaminated by various unwanted artefacts, even though we try to limit their occurrence during the recording session. These artefacts are entangled with the desired brain activity and can have an amplitude up to 100 times that of the brain activity. In most EEG we encounter the following undesired artefacts: ocular, muscular, cardiac, perspiration, line noise, etc. (\cite{Icalabel, Uriguen2015} give more details). Another difficulty that we may encounter during EEG analysis is the volume conduction, i.e. the transmission of electric fields from a primary current source through biological tissue towards the recording electrodes \citep{Olejniczak2006}. Because of volume conduction, unwanted artefacts will impact a broader region and therefore will contaminate more electrodes. In addition, we lose the ability to study a single source or brain region of interest; information is diluted and a signal recorded at one electrode is a combination of all the electrical activities present elsewhere \citep{Uriguen2015}.\\

\noindent Parkinson's disease diagnosis using EEG has been studied in several works. \cite{Cavanagh2018} uses a selection of Fourier transform coefficients to achieve a maximum accuracy of $82\,\%$. It is to be noted that in our study we use the same data as the former. \cite{Oh2020} proposes a fully automated approach based on a \mbox{1-Dimensional} Convolutional Neural Network (1-D~CNN). The model directly classifies the temporal EEG epochs achieving an accuracy of $88.2\,\%$. To perform the diagnosis, \cite{Bhurane2019} relies on correlation coefficients calculated between channels as well as the coefficients of an AR model identified on the EEG to yield a presumable accuracy of $99.1\,\%$. \cite{Yuvaraj2018} uses high-order spectra to perform the diagnosis by extracting thirteen features from the EEG frequency spectrum, he achieved a presumable accuracy of $99.25\,\%$. \cite{Han2013} uses the coefficients of an AR model and the wavelet packet entropy to analyse and investigate whether there is a difference between the parkinsonians and the healthy individuals with no attempt to separate the subjects. Finally, \cite{Liu2017} utilises entropy-based features of 10 channels and a three-way decision model to obtain a classification accuracy of $92.9\,\%$. This last study would have been more relevant if the author addressed the problem of unbalanced data-set. We note that the majority of studies are based only on the frequency features of the EEG and that few studies focus on the temporal features while the two domains should complement each other. Only a few of the features used are explainable and we can understand their design basis to derive conclusions for future work.\\

\noindent We strongly believe that some of the above mentioned methods \citep{Cavanagh2018, Oh2020, Bhurane2019, Yuvaraj2018} are subject to data leakage problems. Data leakage is defined as the use of information in the model training process that is not supposed to be available at the time of prediction \citep{Kaufman2012}. This would not be possible in a real life scenario, where we receive new samples of unlabelled data that we need to categorise. This data leakage will bias the evaluation of the model, which will perform better on the available data used for training, but will perform poorly on the new data. The first type of data leakage that some of the proposed methods suffer from is group leakage, where correlated data from the same subject are present in both the training and the test sets \citep{Ayotte2021}. In this case, and using limited amounts of data, a complex model such as the 1-D CNN can even identify the subject's signature. The second type of data leakage is the fact of optimising hyper-parameters and perform feature selection directly on the test-set (absence of a validation set) \citep{Kaufman2012}.\\

\noindent The aim of this paper is to propose a method for PD diagnosis using EEG signals recorded during a \mbox{3-oddball} auditory task. The data at our disposal are composed of \mbox{$N\!=\!50$} subjects, of which 25 patients suffering from Parkinson's disease. Our main focus is not to have the highest accuracy at any cost, but rather to develop a valid method with minimal bias. We aim to identify new biomarkers that go beyond traditional EEG statistics and spectral content as found in the literature, but instead consider the combination of frequency content, dynamics, and temporal aspects of the EEG.\\

\noindent The proposed method has several notable advantages:
\begin{itemize}[noitemsep,topsep=3pt,parsep=3pt]
\item It is inspired by the current understanding of how cognitive processes and brain works.
\item It involves explainable features that may lead to future studies.
\item It has the potential to work for the early and late stage disease diagnosis.
\item It utilizes a simple and interpretable model with low computational demands.
\item It has been rigorously constructed to avoid data leakage issues.
\item It has been validated on a publicly available database, with transparent and open access implementation code and clearly described execution steps.
\end{itemize}\ \\[-12pt]

\noindent In the present paper, the data we used and the pre-processings we applied are presented in Section~\ref{Dataset}. In Section~\ref{Method}, the basic concepts of our method and its resemblance to the mechanism underlying cognitive processes are described, furthermore, how the proposed idea can be put into practice from a mathematical point of view is also described in this section. In Section~\ref{Results}, the results obtained are presented along with the various validity and robustness tests, moreover, an evaluation in a more constrained settings is provided. Finally, the conclusion of our work is outlined in Section~\ref{conclusion}.

\section{Dataset} \label{Dataset}
\noindent First of all it is to note that our method is agnostic to the dataset selection as long as it contains EEG data. Moreover it is straightforwardly applicable if the EEG was recorded during a 3-oddball auditory experiment.
Several EEG datasets dealing with the diagnosis of PD exist, we examined these before selecting the one on which we can evaluate and test our method. As we want a significantly large number of patients as well as a large amount of data, the following dataset: \url{http://predict.cs.unm.edu} \textit{(ID: d001)} \citep{Cavanagh2017} was chosen. For clarity and reproducibility, we tested our method on a publicly available dataset and made our implementation code accessible to the public through the link: \url{https://github.com/HoussemMEG/SDF_PD}. Additionally, an explanatory animation is included.

\subsection{Data-set description} \label{Data-set description}
\noindent The experimental EEG data available was recorded from \mbox{$N\!=\!50$} participants, $25$ of whom were suffering from PD and an equal number of sex and age matched participants serving as a control group (CTL). The PD group were subject to the same experiment twice, once on-medication and the other time off-medication. In this document, we only consider the off-medication sessions as they showed a noticeable separability from the CTL group in comparison to the on-medication sessions.\\

\noindent The PD group were subject to a Unified Parkinson's Disease Rating Scale (UPDRS) assessing the severity of their disease which was scored by neurologists, the mean UPDRS score is ($24.80 \pm 8.66$). All participants underwent a Mini Mental State Exam (MMSE), and all obtained a scored above 26 (PD: $28.68 \pm 1.03$, CTL: $28.76 \pm 1.05$) confirming their ability to comprehend the task they would be subjected to. Complete details and informations regarding the subjects and the experimental procedure can be found in \citep{Cavanagh2018}.\\

\noindent The experiment consisted of a 3-Oddball auditory task, during which the subjects were presented with a series of 200 repetitive auditory stimuli (trials) infrequently interrupted by a deviant stimulus. Three types of stimuli can be distinguished: 
\begin{enumerate}[noitemsep,topsep=6pt,parsep=2.5pt,partopsep=5pt]
\item \textit{Standard} ($70$ \% of the trials).
\item \textit{Target} ($15$ \% of the trials).
\item \textit{Novel} / \textit{Distractor} ($15$ \% of the trials).
\end{enumerate}\ \\[-14pt]

\begin{figure}[h]
\centering
\includegraphics{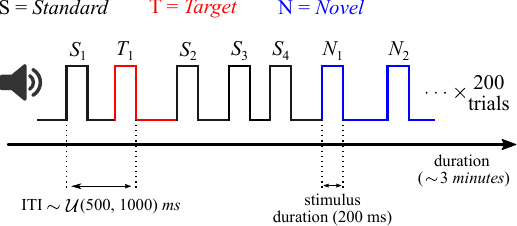}
\caption{Example of a sequence of auditory stimuli.} \label{fig:auditory stim example}
\vspace{-4pt}
\end{figure}

\noindent During this task, the subjects had to count the number of \textit{target} stimuli they had heard throughout the whole experiment. The auditory stimuli were presented for a period of $200$ ms and were separated by a random Inter-Trial Interval (ITI) drawn from a uniform distribution of ($500$ --- $1000$) ms preventing subjects habituation and anticipation. Figure.\ref{fig:auditory stim example} draws an example of an auditory stimuli sequence.
\subsection{Data analysis and pre-processing} \label{Data analysis and pre-processing}
\noindent Throughout the experiment, the EEG signal was continuously recorded at a sampling rate of $f_s\!=\!500$ Hz by the mean of $64$ electrodes (channels). Very ventral temporal sites were removed by \citep{Cavanagh2018} as they tend to be unreliable, leaving at the end $60$ channels. The data were then re-referenced to an average reference.\\ 

\noindent As mentioned in the introduction part, EEG signals are known to be very noisy and present many practical difficulties. Indeed, the coveted brain activity is of a low amplitude and is often drowned out by ambient noise, making the pre-processing stage mandatory. Despite the intrinsic complexity of EEGs and their noise content, the pre-processing steps we have applied are very mild due to the fact that our method is robust to noise. Firstly, to separate and disentangle the unwanted, high-amplitude ocular activity from the coveted cerebral activity, we conducted an Independent Component Analysis (ICA) on the data \citep{Tharwat2018}. We analyzed each independent component (IC) of each subject individually, the ICs that contained eye blinking were removed by projection following the guidelines and recommendation of \citep{Icalabel} and \citep{Cavanagh2018}. Secondly, the data were then bandpass filtered using a Hamming window, attenuating the frequencies outside the \mbox{($1$ --- $30$) Hz} interval. This frequency interval was selected because many studies take 20 or 30 Hz as the upper filtering limit \citep{Starkstein1989}. We have taken the widest interval knowing that our method remains valid even if we widen this interval further.\\

\begin{figure*}[!htb]
\centering
\includegraphics{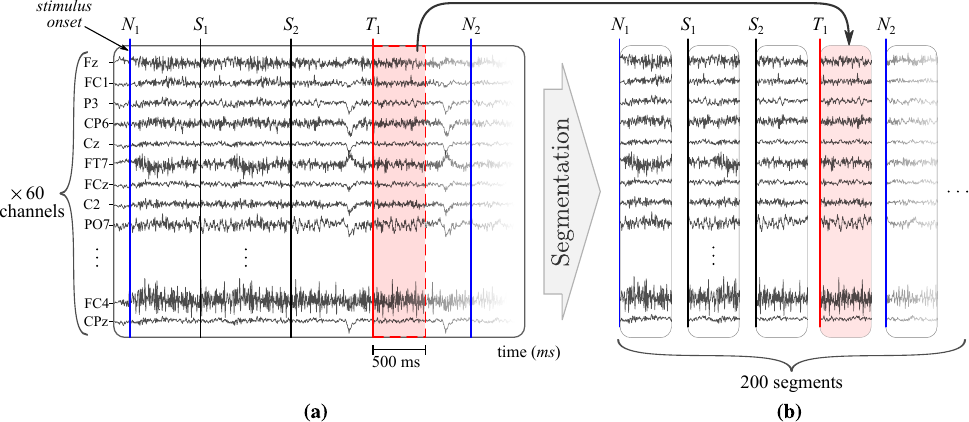}
\caption{EEG segmentation. \textbf{(a)} Raw and continuous 60-channel EEG, recorded for a duration of ${\sim}3$ minutes. Stimuli are time-locked to their arrival instant. Segments of 500 ms were taken starting from the stimulus onset. \textbf{(b)} The result of the segmentation: a 60-channel signal divided into 200 time windows, each time window corresponds to one stimulus response.} \label{fig:segmentation}
\end{figure*}

\begin{figure*}[!htb]
\centering
\includegraphics{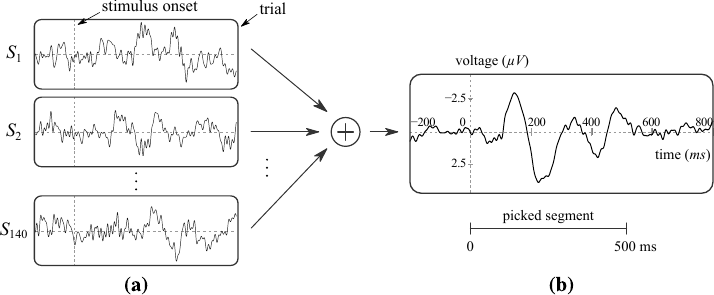}
\caption{Example of ERP creation process for the \textit{Standard} stimulus on one specified channel. \textbf{(a)} Vertical averaging of the corresponding \mbox{time-locked} EEG segments. \textbf{(b)} Result of the vertical averaging process. The signals length used in this example is only demonstrative, the true window length used in our method is indicated at the bottom.} \label{fig:ERP creation}
\end{figure*}

\noindent Time windows (segments) starting from stimulus onset \mbox{($0$ ms)} up to \mbox{($+500$ ms)} post-stimulus were formed, resulting in $200$ time-locked segments, one for each stimulus (see Fig.\ref{fig:segmentation} for a more detailed graphical representation). An event-related potential (ERP) was also calculated separately for each stimulus type by vertically averaging all the signal segments corresponding to the same stimulus type and channel (see Fig.\ref{fig:ERP creation}) \citep{luck2005an}. The aim of this step is to filter the signal and sum up the events occurring at the same time to make them stand out from the ambient noise. Moreover, all the pre-processing steps were performed using MNE (An open-source Python package for exploring, visualizing, and analyzing human neurophysiological data) version $0.24.1$ \citep{Gramfort2013}.

\section{Methodology} \label{Method}
\subsection{Idea and inspiration} \label{Idea and inspiration}
\noindent To process, encode, retrieve and transmit information, biological neuronal networks oscillate \citep{Ward2003, Buzsaki2004}.  The frequencies and timings (phase) of these oscillations are important as they are at the basis of the mechanism underlying cognitive processes \citep{Basar2001, Fries2005}. As suggested by many studies, oscillation frequencies are task dependent \citep{Ward2003}. The oscillation timing is of a great importance since it carries the information about the neuronal dynamics and it is also what makes neuronal synchronization possible. This latter plays a crucial role in cognitive processes \citep{Ward2003, Buzsaki2004}. It should be noted that the EEG mainly measures the electrical potential of a group of neurons oscillating in synchrony. Due to volume conduction, the recorded EEG is the result from the combined activity of different electrical sources distributed in several regions of the brain. The mechanism of synchronization and desynchronization of a group of neurons suggest that the rhythms contributing to the EEG occur in a \textit{pulsatory} \mbox{manner \citep{Olejniczak2006}}.\\
\newpage
\noindent We assume that the ERP associated to a stimulus can be decomposed into two types of rhythms:
\begin{enumerate}[noitemsep,topsep=6pt,parsep=2.5pt,partopsep=5pt]
\item An ongoing activity: it refers to the oscillatory state of the brain when no stimulus is perceived (spontaneous background activity).
\item An evoked activity: it refers to the brain rhythms in response to a stimulus. It starts around $200$ ms after stimulus onset and then vanishes in a time comprised between $500$ ms --- $700$ ms.
\end{enumerate}\ \\[-14pt]

\noindent By having a set of $m$ harmonic oscillators (pendulums) with distinct natural pulsations $\{\omega_1,\, \omega_2,\, \hdots,\, \omega_m\}$, we can represent our signal as a combination of these oscillators where each one of them contributes with its own amplitude according to the Fourier decomposition. However, the temporal information about when an oscillator is turned on or off cannot be recovered by the Fourier transform. Since several rhythms can coexist at the same time, at the same or at different places and moreover interact with each other, the temporal information that cannot be recovered using the Fourier transform, is of great importance. For a given ERP signal, the design objectives of our method are the following:
\begin{itemize}[noitemsep,topsep=5pt,parsep=2.5pt,partopsep=5pt]
\item Disentangle and separate the contribution of each harmonic oscillator.
\item Indicate which oscillators are involved as well as their amplitude contribution.
\item Indicate when an oscillator is turned on or off.
\item Separate the ongoing activity from the stimulus response.
\end{itemize}
The pulsatory nature of brainwaves guided us towards using the principle of parsimony in our method. A high level of parsimony leads to the use of fewer oscillators and they will be excited less frequently (pulsatory fashion). By adjusting the parsimony level we can control the number of oscillators that can be awakened / used by our method and the frequency of their usage.\\

\noindent Studies have shown that patients with PD tend to be significantly slower than healthy individuals (latency observed in auditory ERP) \citep{Hansch1982,Ferrazoli2022}. They also tend to have lower ERP amplitude over certain brain regions compared to CTL individuals \citep{Polich2007, Ferrazoli2022}. However, even if this difference is observed between PD and CTL groups, the last two biomarkers extracted directly from the ERP do not permit the discrimination between the two groups with good separability and accuracy. This is mainly due to the large variance in latency and amplitude within the two groups, as well as the high sensitivity of these two indicators to the disease duration and severity \citep{Ferrazoli2022}. We believe that the intra-group variance is due to the high noise content of the EEG, which makes the extracted measures very sensitive and variable among patients of the same group.\\

\noindent We believe that if we change our point of view on the data, and we do not look directly at the raw EEG but rather try to break it down into its basic components could lead to the development of new biomarkers. By basic components we are referring to oscillators that are turned on and off at a specific timing. Among these new biomarkers, we can find: the latency and activation time of the oscillators, the amplitude or energy content of these oscillators, etc. We believe that these new biomarkers can significantly separate, with good accuracy and robustness\footnote{By robustness we are referring to: consistency, statistical validity, the fact that changing a hyperparameter does not change the result, and simplicity (low variance).}, the PD group from the CTL group.

\subsection{Sparse Dynamical Features} \label{Sparse Dynamical Features}
\noindent In order to put the idea described in the previous section into practice, we model the EEG response by the mean of a battery of pendulums with distinct angular frequencies. These pendulums are turned on and off at a specific time with the appropriate excitation amplitude, so that our model's prediction $\hat{y}(k)$ fits the EEG signal of interest\footnote{Not the entire EEG signal, but a single signal over a specified channel and window. It can also be, as we have used, an ERP.} $y(k)$.\\

\noindent Consider $m$ oscillators with distinct angular frequencies  $\{\omega_1,\, \omega_2,\, \hdots,\, \omega_m\}$. The system combining the $m$ decoupled harmonic oscillators can be described by the following discrete state space representation given a sampling period $T_s = 1/f_s$:
\begin{equation}
\label{eqn: discrete system}
\mathlarger{\Sigma}\!: \left\{\!
	\begin{array}{l}
		x_{k+1} = Ax_k + Bu_k\\[2.5pt]
		\hat{y}_k = Cx_k
    \end{array}\!\!; 
\right. \quad x_k\!\in\!\mathbb{R}^{2m}, \:\: u_k\!\in\!\mathbb{R}^{m}, \:\: \hat{y}_k\!\in\!\mathbb{R}
\end{equation}
\noindent where $x_k$ is the state vector, $u_k$ is the input vector and $\hat{y}_k$ is the output of the system $\mathlarger{\Sigma}$, and:
\begin{equation}
\notag
A\!=\!\text{diag}(a_1,\, \hdots,\, a_m),\quad
B\!=\!\text{diag}(b_1,\, \hdots,\, b_m),\quad
\setlength\arraycolsep{3.5pt}
\renewcommand*{\arraystretch}{0.8}
C\!=\! 
\begin{pmatrix}
c_1 & \cdots & c_m
\end{pmatrix}
\end{equation}
It is to note that $\text{diag}(a_1,\, \hdots,\, a_m)$ is a block diagonal matrix with the elements $\{a_1,\, \hdots,\, a_m \}$ being on the main diagonal and zero entries elsewhere.
\begingroup
\begin{flalign}
\label{eqn: one oscillator}
&
\text{More precisely:}
&&
a_i = 
\begin{pmatrix}
1 & T_s\\
-T_s \omega_i^2& 1\,
\end{pmatrix}, \quad  
b_i = 
\begin{pmatrix}
0\\
T_s
\end{pmatrix}
, \quad 
c_i = 
\begin{pmatrix}
f_s \,\omega_i & 0
\end{pmatrix} &
\end{flalign}
\endgroup
\noindent are the matrices of a single harmonic oscillator: $a_i$ contains the pendulum dynamics, $b_i$ is the input matrix, $c_i$ is the output matrix.\\

\noindent The $m$ oscillators are merged throughout the matrices $A$, $B$ and $C$ such that the $m$ oscillators remains decoupled but all their contribution is summed to form the signal $\hat{y}(k)$. In the control theory field, the oscillators described by the matrices \eqref{eqn: one oscillator} are called modes. The input vector $u_k$ contains the excitation force of all the $m$ modes at instant $k$, \mbox{$u_k = \pig(\smallU_{1}(k)\;\: \smallU_2(k) \;\: \hdots \;\: \smallU_m(k) \pig)^T$}. The form of the matrix $b_i$ has been chosen such that the $i^{th}$ pendulum is controllable by its corresponding control input $\smallU_i(k)$.\\

\noindent The role of the scaling term $\omega_i$ present in the matrix $c_i$ is to ensure that an oscillator at rest, excited by a force $\smallU_i(k)=z$, starts oscillating at time instant $k$ with the same amplitude as the excitation, i.e. $z$. This gives a physical meaning to the values of $u_k$ as they will be directly proportional to the oscillations amplitude of the EEG / ERP. This also opens up the possibility of comparison between the $\smallU_i$ values since they are now on the same scale. We can now compare the modes in terms of their amplitude contribution, but also compare their energy, activity duration, starting time, number of times they have been excited, etc. From this representation it can be noted that the analysis of the excitation forces $u$ can be conducted on a single specific oscillator or a desired set of oscillators. \\

\noindent The system \eqref{eqn: discrete system} can be then rewritten in the explicit form as:
\begin{align}
x_k = A^kx_0 + \sum_{i=0}^{k-1}\,A^iB\, u_{(k-1-i)}, \quad k = 1,\, 2,\, \hdots \\
\hat{y}_k = C A^kx_0 + \sum_{i=0}^{k-1}\, C A^i B\, u_{(k-1-i)}, \quad  k = 1,\, 2,\, \hdots \label{eqn: explicit form}
\end{align}
where $x_0$ represents the initial oscillating state of our model (initial position and velocity of each pendulum). If no external force is applied on these pendulums as in the case of the non-forced regime, i.e. $u_k\!=\!0$ for $k\!=\!1,\, 2,\, \hdots$ , the pendulums will keep swinging in the same manner. Therefore, the information about the background activity of the brain in the case where no stimulus is perceived is carried by $x_0$.\\

\noindent For a given signal of interest $Y$ of length $L$, which we want our model $\mathlarger{\Sigma}$ to fit, we define:
\begin{equation}
\label{eqn: signal vectors}
\begingroup
\renewcommand*{\arraystretch}{1.2}
Y = \begin{pmatrix}
y_0\\
y_1\\
\vdots\\
\,y_{L-1}\,
\end{pmatrix} \in \mathbb{R}^L \qquad
\hat{Y} = \begin{pmatrix}
\hat{y}_0\\
\hat{y}_1\\
\vdots\\
\,\hat{y}_{L-1}\,
\end{pmatrix} \in \mathbb{R}^L \qquad U = \begin{pmatrix}
u_0\\
u_1\\
\vdots\\
\,u_{L-2}\,
\end{pmatrix} \in \mathbb{R}^{m(L-1)}
\endgroup
\end{equation}
where $\hat{Y}$ is the predicted output signal of the system $\mathlarger{\Sigma}$ and $U$ is the control sequence (how our pendulums are swung and excited over time). By forming the vector $U$ in this manner, the information of which of the $m$ oscillators is activated is embedded in the element $u_k$. The temporal information of when the excitation arrives is indicated by the subscript $k$.\\

\noindent Given \eqref{eqn: signal vectors}, the equation \eqref{eqn: explicit form} can be written in a matrix form as:
\begin{equation}
\label{eqn: linear phi equation}
\begin{pmatrix}
\setlength\arraycolsep{0pt}
\phi_1 & \phi_2
\end{pmatrix} \cdot
\begingroup
\renewcommand*{\arraystretch}{1.2}
\begin{pmatrix}
x_0\\
\,U\,
\end{pmatrix}
\,=\, \phi \cdot \beta \,=\, \hat{Y}
\endgroup
\end{equation}
\vspace{-8pt}
\begingroup
\renewcommand*{\arraystretch}{1.1}
\begin{flalign}
\notag &
\text{Where:}
&&
\phi_1 = \begin{pmatrix}
C\\
CA\\
\vdots\\
\:CA^{L-1}\:
\end{pmatrix} \,\in \mathbb{R}^{L\, \times\, 2m}, \qquad 
\phi_2 = \begin{pmatrix}
0 & 0 & \hdots & 0\\
CB& 0 &  \hdots & 0\\
CAB & CB &  \hdots & 0\\
\vdots & \vdots & \ddots & \vdots\\
\:CA^{L-2}B & CA^{L-3}B & \hdots & CB\:
\end{pmatrix}\, \in \mathbb{R}^{L\, \times\, m(L-1)}
&
\end{flalign}
\endgroup

\subsubsection{Virtual Modal Stimuli generation}

\noindent A simple and cheap primary solution that finds $\beta$ and  minimizes the squared error $\epsilon = \norm{\, Y - \phi\, \beta \,}^2_2$ from equation \eqref{eqn: linear phi equation} is given by the least square method \citep[chap.~2]{Hastie2009-ag}. Despite its simplicity, it does not meet our expectations due to the fact that the resulting $\beta$ will be full (not sparse). As we mentioned in the Section \ref{Idea and inspiration}, we want to reproduce the pulsating nature of the brain rhythms, which should be achieved by a sparse $\beta$. Moreover, it should be noted that the minimization of the squared error will not necessarily yield the lowest classification error and the highest classification robustness.\\
\noindent Several subset selection and shrinkage methods exist (for more details see \cite[chap.~3]{Hastie2009-ag} for example). To induce sparsity in the $\beta$ solution, we have chosen Lasso-LARS algorithm (Least Absolute Shrinkage and Selection Operator --- Least-Angle Regression) \citep{Efron2004}. Briefly, instead of minimizing only the squared error, Lasso-LARS penalizes the number of non-zero entries of $\beta$ through a weighted $l_1$ norm. The corresponding cost function is:
\begin{equation}
\label{eqn: cost function}
\hat{\beta} = \argmin_{\beta}\; \dfrac{1}{2}\, \norm[\big]{\,Y - \beta_0 - \phi\, \beta\,}_2^2 + \alpha\, \norm[\big]{\,\beta\,}_1
\end{equation}


\noindent where $\beta_0 \in \mathbb{R}$ is the intercept\footnote{Mean value of the response variable when all of the predictor variables in the model are equal to zero. In this case it is equal to $\bar{Y}$.} term, $\hat{\beta}$ is the estimated $\beta$ solution and $\alpha$ is a weighting constant (more details will be given below).\\
\noindent We have opted for the Lasso-LARS method over others for its inherent ability to compute the full solution path as $\alpha$ varies. Moreover, it is a very efficient algorithm to compute the solution of the Lasso problem, particularly when $\text{dim}(\beta)\mkern-1.2mu \gg \mkern-1.2mu L$, which corresponds to our case. We used the Lasso-LARS implementation of scikit-learn\footnote{Open source python library for data analysis and machine learning.} version 1.0.2 \citep{scikit-learn}. Throughout the rest of this paper. the vector $\beta$ in equation \eqref{eqn: linear phi equation} is named VMS for Virtual Modal Stimuli.


\begin{figure*}[!htb] 
\centering
\includegraphics{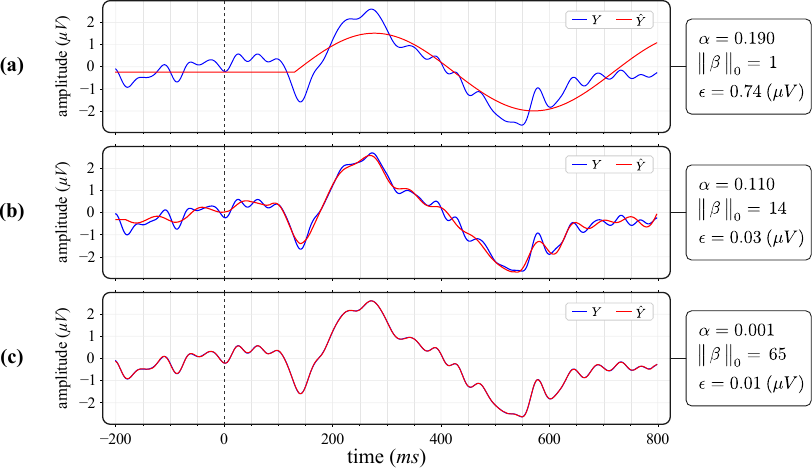}
\caption{Effect of $\alpha$ values on the predicted signal $\hat{Y}$ and on the number of VMS mobilised by our model. $\norm{\, \beta}_0$ indicates the number of non-zero entries of $\beta$. \textbf{(a)} Shows the activation of a single mode, selected with the appropriate frequency and excited at the right time. \textbf{(b)} Utilisation of more modes and its effect on the $\hat{Y}$ prediction. \textbf{(c)} Shows the ability of our model to fit the $Y$ signal by exciting more modes and more frequently (65~excitations). \label{fig:alpha varying}}
\end{figure*} 
\noindent In our case, the best solution that yields the best result does not necessarily have to fit the signal $Y$ perfectly as there is a trade-off between fitting the signal $Y$ tightly and risking to capture the noise as well, and not fitting the signal $Y$ tightly at the risk of not capturing the important information. We are looking for something in between, to do so, we adjust the value of $\alpha$ to control the sparsity of the VMS and thus controlling the fitting level of our model (see Fig.\ref{fig:alpha varying}). For a large value of $\alpha$, the sparsity is high and only a few modes are excited, while for a small $\alpha$ value, the sparsity is low, so more modes are utilised by the model.\\[-1pt]
\noindent At the initial stage of the algorithm no modes are excited, then, at each iteration the Lasso-LARS starts to excite and awaken modes. The $\alpha$ value at this step corresponds to a sparsity level of $100$ \% and is calculated on the basis of the correlations (see \cite{Hastie2009-ag} for more details). After several iterations, the algorithm will utilise sufficient number of modes to reach the final $\alpha_{f}$ value provided to the algorithm. This low $\alpha_{f}$ corresponds to a low sparsity level of $0$ \%. Between the initial value and the final value of $\alpha$, the Lasso-LARS generates the entire solution path of the equation \eqref{eqn: cost function} for decreasing level of sparsity. Path length depends on the specified $\alpha_f$, signal complexity and the $\phi$ matrix (number of oscillators $m$ that we included in the model).\\

\noindent The value of the hyper-parameter $\alpha_{f}$ that yields the best results is not known beforehand. Thus, we selected several solutions along that path corresponding to several levels of sparsity by slicing the interval $\big(100$ --- $0\big)$ \% linearly with $level\, \%$ steps. Finally, we evaluated the classification results for each of these $\beta$ solutions with varying levels of parsimony. It should be noted that due to this selection, the exact value of $\alpha_f$ is not as important as its magnitude.\\
\noindent The Lasso-LARS algorithm is stated in terms of correlations, so if the input features (columns of $\phi$) are standardised, the algorithm will run faster as it will utilise inner-products. The standardisation serves also to mitigate the effect of the differing magnitude of the input features on the correlations, therefore the entries of $\beta$ have an equal chance of being selected a priori.\\
\noindent Instead of standardising the $\phi$ matrix in the conventional way, we have made a small modification. Let's denote $\underline{\phi_{1}}$ and $\underline{\phi_{2}}$ as the vector-wise standardised version of the matrices $\phi_1$ and $\phi_2$ respectively. We have introduced a weighting constant $w$ so that the final standardised $\underline{\phi}$ matrix is created by concatenation as follows:
\begin{equation}
\underline{\phi}\,=\, \begin{pmatrix}
w\,\underline{\phi_1} & (1-w)\,\underline{\phi_2}\mkern+2mu
\label{eqn: merged matrix}
\end{pmatrix}
\end{equation}

\begin{figure*}[!htb]
\centering
\includegraphics[scale=0.98]{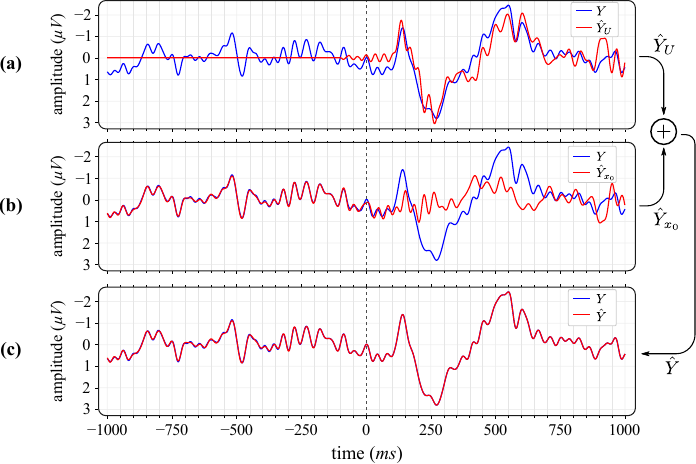}
\caption{Example illustrating how a signal of interest $Y$ (in blue) is fitted by our model \eqref{eqn: discrete system} (in red). For this example, we have used $\alpha_f\!=\!8\! \times \!10^{-4}$ and $w\!=\!0.55$ \textbf{(a)} Estimated contribution of the forced regime alone $\hat{Y}_{U}$. As shown in the sub-figure, the modes are triggered near the stimulus onset and the shape of this response matches the shape of the ERP. \textbf{(b)} Estimated background rhythm $\hat{Y}_{x_0}$, the oscillations match the ERP response prior to stimulus arrival, and keep oscillating in the same fashion. \textbf{(c)} The contribution of (a) and (b) are summed to give the fitted signal $\hat{Y}$.} \label{fig:yseparation}
\end{figure*}

\noindent This step is done in order to manage the trade-off between a priori likelihood and favoring/guiding the choice of modes that are chosen first (either those of $x_0$ or $U$). Therefore, the factor $w$ serves to adjust whether we put more emphasis into the ongoing activity or the evoked activity. A value of $w=1$ will place all the predicted VMS only in $x_0$ while a value of $w=0$ will place all the VMS in $U$. It also allows to balance between the number of entries that will be distributed either in $x_0$ or in $U$ of the final solution. This variation was made to favour adjusting for the $x_0$ features first since they affect the whole prediction unlike the $U$ which are active only on a specific small time span (see Fig.\ref{fig:yseparation}). This step will make the algorithm more stable and give better results.\\

\noindent The Lasso-LARS algorithm is mainly used to induce sparsity in the solution of $\beta$. However, the main objective of the algorithm is to minimize the cost function \eqref{eqn: cost function} and to find its estimated optimal solution. As the Lasso-LARS algorithm is shaped, the final resulting $\beta$ is not completely fitted, i.e. to minimise the cost function \eqref{eqn: cost function} $\beta$ must be small in magnitude so that $\norm{\, \beta\,}_1$ is small. As stated previously, we are only interested in finding the few most important entries of $\beta$ that are used in fitting as much as possible the signal $Y$. Once $\beta$ is computed, we will only pick its non-zero entries and used them to fit the $Y$ signal using the Least Square algorithm. This reduces the estimation error $\epsilon$ without increasing the number of modes and their frequency of use. The temporal information of when a mode is excited is also preserved.\\

\noindent From the equation \eqref{eqn: linear phi equation}, we can observe that the predicted signal $\hat{Y}$ can be decomposed into two parts: (1) an ongoing activity contribution $\hat{Y}_{x_0}$ (heavily dependent on the environment and not necessarily on the patient) and (2) an evoked activity contribution $\hat{Y}_U$. On Figure~\ref{fig:yseparation} we can visualize how the signal $Y$ is decomposed into a background rhythm and a stimulus response. The $Y$ signal used in this example is an ERP associated with \textit{standard} stimuli over the \mbox{channel $C_z$}.
\begin{equation}
\label{eqn: yx0 and yu}
\hat{Y} = \underbrace{\phi_1\, x_0}_\text{\footnotesize$\hat{Y}_{x_{0}}$} + \underbrace{\phi_2\, U}_\text{\footnotesize$\hat{Y}_{U}$}
\end{equation}
\noindent For the rest of the method, we will only consider the forced regime response (the second part of the equation) and therefore only $\hat{Y}_U$. Thus, we will not analyse $\beta$ in its entirety but only the part regarding $U$.\\[-3pt] 

\subsubsection{Features selection}
\begin{figure*}[h]
\centering
\includegraphics[scale=0.95]{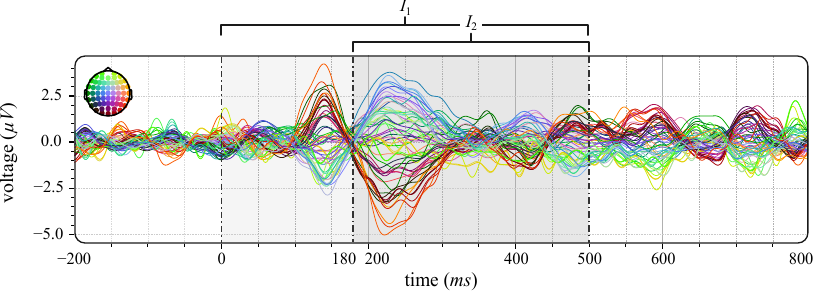}
\vspace{-8pt}
\caption{Features operating intervals, example on a \textit{standard} ERP.} \label{fig:features limits}
\end{figure*}

\noindent Given that we have a limited number of data instances, and for validity purposes of our method, we extracted new features from the generated VMS (concerning $U$ only). The model will rely on these extracted Sparse Dynamical Features (SDF) to classify the subjects as healthy or non-healthy. The advantages of conducting feature extraction~are:
\begin{enumerate}[noitemsep,topsep=4pt,parsep=3pt,partopsep=5pt]
\item Improve the accuracy of the model on the test set.
\item Give physical meaning to our model and facilitate its explicability.
\item Extract only the useful and important information that the model should utilise to perform the classification.
\item Reduce the complexity of the model as we no longer need a very rich model that should address a complex task, but rather have a simple model for a simpler problem.
\item Give new directions for further studies.
\item Enhance model stability, robustness\footnote{Lower accuracy variance under other novel conditions such as the arrival of a new subject.} and increase the confidence placed on the model. The accuracy will not vary much for new, unseen data, which for an application based on a small number of data instances is crucial. Concerning our case, we can even afford to have a test-set size of 50 \% without impacting significantly the classification accuracy (see Table.\ref{tab:result 2-fold}).
\item Reduce the over-fitting, especially as in our case where we have a small number of data instances. If we \mbox{over-fit}, it may make our method impractical and therefore unusable. 
\end{enumerate}
\indent As previously noted in Section \ref{Idea and inspiration}, parkinsonians are mentally slower, with observed latency in the ERP and lower amplitude compared to healthy individuals. We will therefore mould the information contained in $U$ (activation timing and amplitude) to fit these discriminative features. The operating time intervals $I_1$ and $I_2$ of the below SDF are shown in Fig.\ref{fig:features limits}:
\begin{itemize}[noitemsep,topsep=5pt,parsep=0.5pt,partopsep=5pt]
\item $F_1$: the instant of the lowest activation amplitude of all oscillators defined by: $\displaystyle\argmin\Bigl\{U(I_1)\Bigr\}$. 
\item $F_2$: the average excitation forces occurring between $180$ ms and $500$ ms defined by: $mean\Bigl\{U(I_2)\Bigr\}$.
\end{itemize}

\subsection{Classification} \label{Classification}
\noindent The aim of the classification is to use a model, also called classifier, to categorize the input data into different classes. For our case, the input data are the SDF and the output is whether the subject belongs to PD or CTL group. Different classification algorithms exist with different levels of complexity and use cases. The less complex models are often appreciated because of their simplicity of use, their low computational costs and especially for their explainability and the simple interpretation of the results.

\noindent As we only have two features and a straight line or a quadratic curve is sufficient to separate the two classes (see Fig.\ref{fig:decision boundary}), we only tried three simple classification algorithms: Linear Discriminant Analysis (LDA), Quadratic discriminant Analysis (QDA) and Linear Support Vector Machine (linear SVM). It is to be noted that these methods do not have hyper-parameters that we need to tune (more details about these algorithms can be found in~\citep{Hastie2009-ag}).\\
\noindent Among these 3 algorithms, QDA showed the lowest accuracy for our application, while LDA yielded the highest accuracy. Furthermore, the linear SVM had a slightly lower accuracy than the LDA classifier. Throughout the rest of this article, we will only present the results of the LDA classifier.

\subsection{Nested Cross-Validation} \label{Nested Cross-Validation}
\begin{wrapfigure}{r}{0.50\textwidth}
\centering
\vspace{-5pt}
\includegraphics{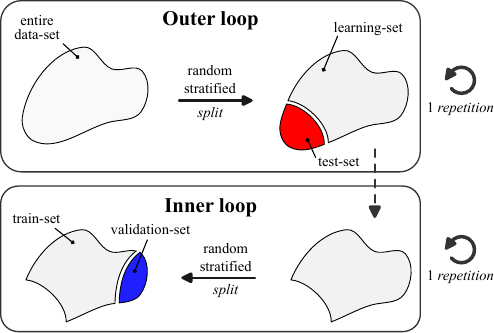}
\vspace{-10pt}
\caption{Nested train-test-validation splitting for a single trial.} \label{fig:Nested train test validation splitting}
\vspace{-3pt}
\end{wrapfigure}
\noindent To evaluate the generalisation performance of our method on new unseen data, we first used the simplest sampling method, namely a Nested Train-Test-Validation split (NTTV) \citep{Molinaro2005}. This procedure consists first of all of splitting the data-set into a holdout test-set and a learning-set, this step takes place in the outer-loop and the splitting is done in a random and stratified manner (keeping the same proportion of the classes as they appear in the initial population if possible). The learning-set is then randomly stratified split into a training-set and a validation-set within the inner-loop (see Fig.\ref{fig:Nested train test validation splitting}). This sampling method allows us to have a validation set in order to select on the latter the best hyper-parameters, but at the cost of reducing the size of the training-set. It is important to recall that we cannot select the hyper-parameters that perform the best on the test-set as this is considered as data-leakage. The latter will lead to a biased model with over-optimistic performances where the model performs very well on the available data but poorly on new, unseen data as is the case in a real-life scenario (see \cite{Varma2006} for a study on this bias).\\

\noindent Firstly, the model is trained on the training-set, then the hyper-parameters\footnote{In our case they are the number of oscillators $m$, their angular frequency $\omega_i$ and most importantly the sparsity level $\alpha$.} are fine tuned and selected on the validation-set. The model is then retrained on the learning-set, using the best hyper-parameter obtained. Finally, we evaluate the trained model performances on the test-set. This assessment, will result in only one performance estimate, which will vary considerably and will depend heavily on the splits (which subject is in which group). One solution to cope with this variance problem is to repeat the splitting procedure several times with different randomization in each repetition, giving several performance estimates, one for each given split. The overall performance will then be the average performance obtained on these different splits. 
However, even if we randomly split the subjects into three sets, some subjects will be chosen much more frequently in one set than in the others. This creates a selection bias that significantly affects the results (the effect is much more pronounced when the dataset is small) \citep{Molinaro2005}. This selection bias is reduced as the number of repeated splits is increased, but at the expense of computational load and time.

\begin{wrapfigure}{l}{0.445\textwidth}
\centering
\includegraphics{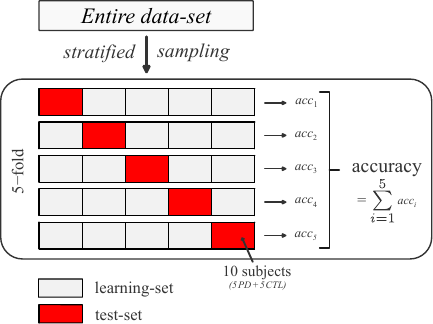}
\vspace{5pt}
\caption{Example of 5-fold Cross Validation.} \label{fig:CV}
\end{wrapfigure}
\noindent K-fold Cross Validation (CV) is another solution to reduce the selection bias that works well on small data sets \citep{Stone1976}. It consists of randomly partitioning the data set into K-folds (K partitions). For K times, we keep a fold that was not previously chosen as a holdout test-set and use the remaining $K\! -\! 1$ folds as a learning-set (see Fig.\ref{fig:CV} for $K\!\!=\!\!5$ example). K may be set to the total number of data instances so that each observation is the holdout once; This procedure is called Leave-One-Out (LOO) Cross-Validation \citep{Molinaro2005}. This is the best way to reduce the selection bias and improve the learning performance of the model, but at the expense of the computational load \citep{Varma2006, Molinaro2005}. In our case, this is practicable because our model is simple and computationally inexpensive, moreover, we are working on a small data~set.\\

\begin{figure}[h]
\centering
\includegraphics{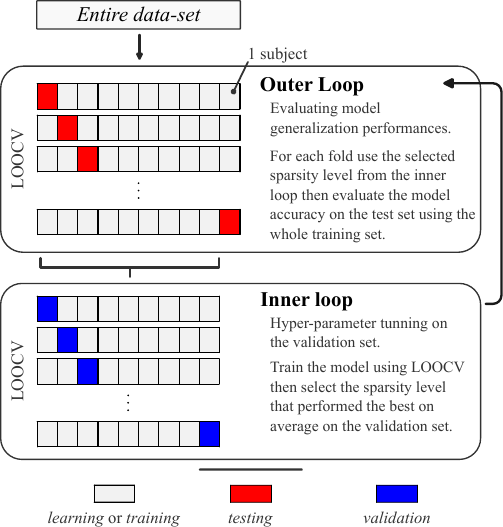}
\caption{Schematic display of the nested cross-validation.} \label{fig:nested CV}
\end{figure}

\noindent To take advantage of the benefits of the two sampling methods presented above (NTTV and LOO) and to compensate for their drawbacks, we combined them to obtain the Nested Leave-One-Out (NLOO) Cross-Validation. This combination enables us to do a hyper-parameters tuning, reduce the selection bias and compensate for the small training-set bias present in NTTV. The NLOO cross-validation procedure is composed of two nested loops: an outer-loop and an inner-loop, in each loop there will be a LOO procedure. The outer-loop splits the entire data-set into a learning-set and a test-set while the inner-loop splits the learning-set into a train-set and a validation-set (for more details, see Fig.\ref{fig:nested CV}).
For the rest of this article, the internal loop always corresponds to a LOO CV procedure. However, with regard to the outer-loop, the case where we have a K-fold CV instead of a LOO CV will be mentioned. Moreover, the accuracies presented in this paper are the average accuracies obtained on the test-set over all the repetitions, the case where the accuracy concerns the train-set or validation-set will also be mentioned.

\section{Results} \label{Results}
\noindent For each subject, we computed the ERPs corresponding to each stimulus type and channel. The window width used ranged from $0$ ms to $500$ ms post-stimuli\footnote{The upper limit of $500$ ms was chosen due to the minimal interval between stimuli $\min(\text{ITI}) = 500$ ms.}, resulting in an ERP of length \mbox{$L = 250$}. The system $\mathlarger{\Sigma}$ was created using $m=40$ pendulums with an evenly spaced angular frequency taken from the interval $2 \pi\!\times\!(1$ to $30$) Hz. The weighting constant $w$ in the equation \eqref{eqn: merged matrix} has been set to $w\!=\!0.55$. We have set this value so as to obtain a separation between the forced and steady-state regime as shown in the Fig.\ref{fig:yseparation} while ensuring that the information is not completely carried by $x_0$ or by $U$ alone. The weighting constant $\alpha_f$ was set to $\alpha_f\! =\! 8\!\times\!10^{-4}$ by visual inspection, so that this value is just small enough to obtain a very good fit. 
We selected the $\beta$ solutions with different sparsity levels by taking the resulting solutions in the interval $\big(100$ --- $0\big)\; \%$ using a regular spacing $level\! =\! 2\, \%$, resulting for each ERP in $50\, \beta$ variants. Finally, once the solution path was computed, we extracted the features $F_1$ and $F_2$ from the generated excitation forces $U$. We used these extracted features to perform the classification using LDA.\\

\noindent The form and complexity of the ERP vary according to the type of stimulus and channel position. This variation is even more pronounced when the channels are far from one another. Our model utilises fewer modes excitations to fit a less complex ERP, while it will utilise more modes excitations to fit a more complex ERP. Therefore, the best working level of sparsity that yields the best classification accuracy varies according to the channel location. Since the best working sparsity level cannot be determined in advance and since we are evaluating our method on all channels and stimuli type, we have left the sparsity level as a hyper-parameter to be tuned during the classification.\\

\noindent As we test a large number of combinations (stimuli type, channels and sparsity level) and have a small number of data-instances, in some cases for a specific condition and at a given sparsity level $\alpha^*_{k\%}$, we observe a good classification accuracy (over $80\,\%$) due to sampling noise \citep{Frost2020}. Fortunately, in this case, we can observe that the surrounding sparsity levels $\alpha_{k-1\%}$ and $\alpha_{k+1\%}$ give much poorer results (an accuracy jump is observed). We have not considered the level of sparsity $\alpha^*_{k\%}$ as a valid result since it is mainly due to luck and is not consistent. We have been careful to take into consideration only the sparsity levels that yielded good classification accuracy while the surrounding levels of sparsity being also good. This accuracy smoothness is expected since the variation in sparsity levels is small. This ensures us that our results are not due to luck, that they are sufficiently consistent and contain valuable information.

\subsection{Stimulus choice} \label{Stimulus choice}
\noindent We started by evaluating the performance of our method for the different stimuli. For each stimulus, we evaluated the accuracy obtained by our model, for all the channels using the NLOO CV. Table.\ref{tab:result per stim type} shows the results obtained. As we are dealing with 60 channels we have only indicated the channel that yielded the highest accuracy (channel*) with its corresponding accuracy.
\begin{table}[h]
\centering
\begin{tabular}{c|ccc}
    \toprule
    Stimuli type & \textit{standard} & \textit{\textbf{target}} & \textit{novel}\\[3pt]
    \midrule
    Channel* & $AF_z$ & $\mathbf{CP_z}$ & $CP_3$\\[1pt]
	Accuracy & $74\,\%$ & $\mathbf{90\,\%}$ & $74\,\%$\\
    \bottomrule
\end{tabular}
\caption{Results of the highest accuracy obtained for each stimulus type with the corresponding channel.}
\label{tab:result per stim type}
\end{table}

\noindent As indicate the table.\ref{tab:result per stim type} the stimulus \textit{target} showed the highest separation using the features $F_1$ and $F_2$. These results, need to be interpreted with caution as other features can lead to different results. Furthermore, the results obtained for the stimuli \textit{standard} and \textit{novel} are not statistically significant. For the rest of this paper we only consider the \textit{target} stimulus.

\subsection{Luck or informative features ?} \label{Luck or informative features ?}
\noindent To test the significance, truthfulness and consistency of the results obtained for the $CP_z$ channel we performed three~tests:
\begin{enumerate}
\item \textit{Permutation test} \citep{Nichols2003}: We define the null hypothesis $H_0$ as: the features $F_1$ and $F_2$ do not allow to differentiate between the PD and CTL group. Under this hypothesis we evaluated the probability that the obtained result was simply due to chance and sampling noise as we have a small data set and we carried out a large number of runs (each per stimuli type, channel and parsimony level) \citep{Frost2020}. After randomly permuting the labels $n\!=\!500$ times, and assessing the best accuracy obtained, we obtain a $p$-value of $\big(p < 0.03\big)$ which indicates and favours the fact that the results are significant. To recall, for statistical significance, the $p$-value should typically be $p < 0.05$.
\item \textit{Parametric consistency}: For this part, unlike the other parts, the parsimony level is fixed (it is no longer a hyper-parameter to be tuned). We used LOOCV for each parsimony level to assess the model's accuracy. Moreover, we set the new spacing $level\!=\!1\,\%$ to have a finer sparsity grid. Figure \ref{fig:accuracy spars variation} shows the accuracies obtained at each sparsity level. The small variation of the accuracy in the parsimony zone that yielded the highest accuracy (see area of interest in Fig.\ref{fig:accuracy spars variation}) may suggest the presence of information and that indeed, by using $F_1$ and $F_2$ the subjects are separable. It should be noted that this area of interest is the area selected in $100\,\%$ of the times during the model selection (in the inner-loop of the NLOO CV). To illustrate how the features $F_1$ and $F_2$ are scattered on the plane, we have plotted these in Fig.\ref{fig:accuracy spars variation} for the sparsity level indicated by a star in Fig.\ref{fig:decision boundary}. In addition, we have also plotted the decision boundary.
\begin{figure*}[!htb]
\centering
\includegraphics{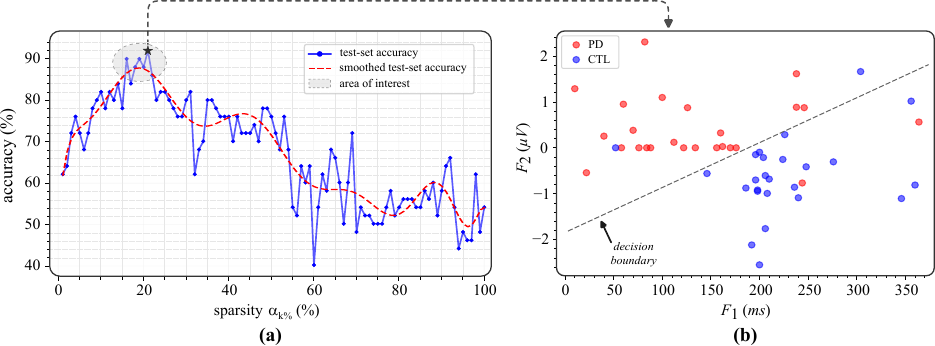}
\caption{\textbf{(a)} Model's accuracy for a varying level of sparsity using LOOCV. \textbf{(b)} Scatter plot of the SDFs with the corresponding decision boundary taken for the sparsity level indicated by a star in (a).} 
\customlabel{fig:accuracy spars variation}{10.a}
\customlabel{fig:decision boundary}{10.b}
\end{figure*}
\item \textit{Spatial consistency}: We have evaluated the spatial evolution of the accuracy for the channels surrounding $CP_z$. If these channels also show good results, this may indeed indicate the presence of information over this region, moreover, it coincides to the suggestions of Fig.7 in \citep{Polich2007}. Figure \ref{fig:result by electrode} shows the accuracies obtained on the test-set after a NLOO CV procedure. We observe that the channels $CP_1$ and $CP_2$ located on the same horizontal line to $CP_z$ showed great results individually, while the other surrounding channels did not yield significant results. We can observe an increasing accuracy as we get closer to $CP_z$, which may suggest the presence of a discriminative information over that region.
\begin{figure}[!htb]
\centering
\includegraphics[scale=0.79]{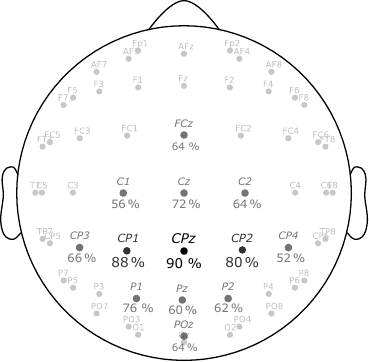}
\vspace{-8pt}
\caption{Model performances over the region surrounding $CP_z$.} \label{fig:result by electrode}
\end{figure}
\end{enumerate}
%

\subsection{Increasing the accuracy through voting} \label{Increasing the accuracy through voting}
\noindent Observing in the Fig.\ref{fig:result by electrode} that the channels $CP_1$, $CP_z$ and $CP_2$ performed well individually, we aggregated the information of these three channels by making them vote on whether the subject is healthy or non-healthy. All the three channels have the same voting power, so if at least two voters agree on a subject's condition, we consider the resulting vote as the output of this new model. Table.\ref{tab:result with vote} reports the accuracies obtained after a NLOO CV procedure. The validation set is non-existent for the voting strategy as we have no parameters to tune.

\begin{table*}[h]
{\renewcommand{\arraystretch}{1.05}
\centering
\begin{tabular}{lcccc}
    \toprule
   \multirow{2}{*}{Accuracy} & \multicolumn{3}{c}{Individual channels} & \multirow{2}{*}{Voting strategy}\\
	 \cmidrule(lr){2-4}   
    & $CP_1$ & $CP_z$ & $CP_2$ & \\
    \midrule
        Learning & $88.0\,\%$ & $90.0\,\%$  & $84.9\,\%$ & $95.4\,\%$ \\
		\hspace{-1.5cm} \makebox(9,0){\textbf{\normalsize(a)}\,\, \rotatebox[origin=c]{90}{NLOO}}\hspace{1.5cm-15pt} Validation & $87.3\,\%$  & $89.9\,\%$  & $84.1\,\%$ & --- \\
		Testing & $88.0\,\%$  & $90.0\,\%$  & $80.0\,\%$ & $\mathbf{94.0\,\%}$ \\
    \bottomrule\\[-9pt] 
    
	\midrule
        Learning $(\pm\, std\,\%)$ 			 & $88.7\,\%\: (\pm\,1.3)$  & $89.7\,\%\: (\pm\,0.7)$  & $86.5\,\%\: (\pm\,1.5)$ & $94.1\,\%\: (\pm\,1.1)$ \\
	\hspace{-1.5cm} \makebox(9,0){\textbf{\normalsize(b)}\,\, \rotatebox[origin=c]{90}{5-fold}}\hspace{1.5cm-15pt} 	Validation $(\pm\, std\,\%)\:\,$ 	 & $87.8\,\%\: (\pm\,1.2)$  & $88.8\,\%\: (\pm\,0.7)$  & $85.7\,\%\: (\pm\,1.4)$ & --- \\
		Testing $(5\!-\!th\,percentile\,\%)$ & $82.7\,\%\: (74.0)$  	& $85.4\,\%\: (79.7)$ 	   & $77.7\,\%\: (70.2)$     & $\mathbf{87.6\,\%}\: (81.9)$ \\
    \bottomrule\\[-9pt] 

    \midrule
        Learning $(\pm\, std\,\%)$ 			 & $89.4\,\%\: (\pm\,2.5)$  & $90.5\,\%\: (\pm\,2.1)$  & $86.6\,\%\: (\pm\,2.1)$ & $94.5\,\%\: (\pm\,2.1)$ \\
	\hspace{-1.5cm} \makebox(9,0){\textbf{\normalsize(c)}\,\, \rotatebox[origin=c]{90}{2-fold}}\hspace{1.5cm-15pt} 	Validation $(\pm\, std\,\%)\:\,$ 	 & $88.3\,\%\: (\pm\,2.3)$  & $88.9\,\%\: (\pm\,2.1)$  & $85.7\,\%\: (\pm\,2.0)$ & --- \\
		Testing $(5\!-\!th\,percentile\,\%)$ & $78.0\,\%\: (67.9)$	 	& $81.7\,\%\: (72.7)$	   & $73.9\,\%\: (65.7)$	 & $\mathbf{83.8\,\%}\: (75.5)$ \\
    \bottomrule 
    
\end{tabular}
\caption{Resulting accuracies obtained for the voting strategy and the three individual channels $CP_1$, $CP_z$ and $CP_2$ using: \textbf{(a)} the nested cross-validation. \textbf{(b)} 5-fold CV ($80\,\%$ learning data and $20\,\%$ test data). \textbf{(c)} 2-fold CV ($50\,\%$ learning data and $50\,\%$ test data).}
\customlabel{tab:result with vote}{II.a} \customlabel{tab:result 5-fold}{II.b} \customlabel{tab:result 2-fold}{II.c} \label{tab:results}
}
\end{table*}

\begin{figure}[!htb]
\centering
\includegraphics[scale=0.9]{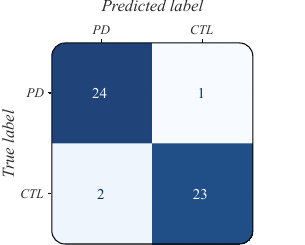}
\caption{Voting strategy confusion matrix.} \label{fig:confusion matrix}
\end{figure}

\noindent Figure \ref{fig:confusion matrix} represents the resulting confusion matrix of the voting strategy. We can observe that the classification is balanced, and the model mislabels the both classes almost equally.
\ \\[-5.5pt]

\noindent Regarding the early diagnosis, as presented in Section \ref{introduction}, in most cases patients suffer from non-motor symptoms before motor symptoms develop. The onset of the latter is slow and in some cases can take decades. The main difficulty of early clinical diagnosis is that it is mainly based on the manifestation of the motor symptoms that have not yet developed sufficiently. To recall, the True Positive (TP) rate\footnote{Ratio of PD patients identified correctly.} of clinical diagnosis is about $26\,\%$ (9 of 34) for a duration since the first diagnosis $<\!3$ years, and of $53\,\%$ (8 of 15) for a duration $<\!5$ years \citep{Adler2014}. Now regarding our data-set, we have 12 patients who have been diagnosed for $<\!3$ years and 2 patients for $<\!5$ years. For these 14 patients, we obtain a $\text{TP} = 100\,\%$ strongly suggesting the utility of our method and its ability to work for early diagnosis cases.

\subsection{Results with less learning data} \label{Results with less learning data}
\noindent For this part, we used K-fold CV instead of the usual LOOCV to evaluate the model performances\footnote{This only concerns the outer-loop, we kept LOOCV for the inner-loop.} under the constraint of having fewer learning data. This step also gives an indication about the ability of our model to generalise to new, unseen data. We believe that if the model trained only by using half of the data performs almost the same as the model trained on the entire data-set it may suggest that the model trained on the entire data-set may obtain similar results on a data-set that are double the size of ours (Under the assumptions that we have no covariate shift, for more details see \cite{Quinonero-Candela2022-ot}).\\

\noindent Table \ref{tab:result 5-fold} reports the accuracies obtained using a 5-fold CV procedure where the learning-set is about $80\,\%$ of the entire data-set size. Table.\ref{tab:result 2-fold} reports the accuracies obtained using a 2-fold CV procedure where the learning-set is about $50\,\%$ of the data-set size. For both 5-fold and 2-fold CV the random partitioning were repeated $30\,000$ times.\\
From the tables \ref{tab:result with vote}, \ref{tab:result 5-fold}, and \ref{tab:result 2-fold} we can observe that the accuracies obtained on the learning and validation sets have not changed significantly, but their variance increases as the learning-set gets smaller. Regarding the test accuracy, it decreased for the case of individual channels and for the voting strategy as the learning-set gets smaller which is expected as less information is present in the learning-set. However, the decrease in accuracy is not substantial, which demonstrates the robustness of our approach, which is mainly due to the utilization of only two well-separable features.\\

\begin{figure*}[!htb]
\centering
\includegraphics{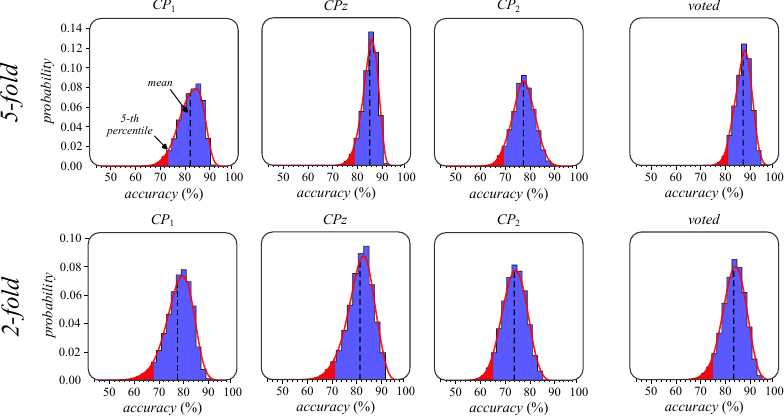}
\caption{Resulting test accuracies obtained for the $30\,000$ random partitioning.} \label{fig:densities}
\end{figure*}

\noindent The test accuracy of the $30\,000$ repetitions is shown in Figure \ref{fig:densities} with the $5\!-\!th$ percentile marked. This indicates that the accuracy of the model is above the indicated value in $95\,\%$ of the cases.

\section{Conclusion} \label{conclusion}
\noindent In this paper we presented a new approach for the diagnosis of Parkinson's disease using EEG signals. We evaluated our method on a public dataset containing $N\!=\!50$ \mbox{subjects of} which 25 have Parkinson's disease and the remaining individuals serve as a control group. The proposed method is inspired by the functioning of the brain and combines the frequency content, the dynamics and the temporal aspect of the EEG. Our approach is based on features called Sparse Dynamical Features (SDF) that are extracted from the ERP signal. The evaluation of our model was carried out with attention to induce the least bias possible so that we do not have an overly optimistic model that only works on our data base but rather a simple model with a strong generalisation capability that can be used on new data. We were able to separate the healthy individuals from the unhealthy ones with an accuracy of $94\,\%$ using only two features by making the models obtained for the channels $CP_1$, $CP_z$ and $CP_2$ vote. We have carried out several tests to verify the validity of our approach. The proposed new biomarkers also work for the case where clinicians face the most problems i.e. the early diagnosis of the disease.

\section*{Acknowledgement}
\noindent This work has been partially supported by MIAI@Grenoble Alpes (ANR-19-P3IA-0003).

\section*{Conflict of interest}
\noindent None of the authors have potential conflicts of interest to be disclosed.





\bibliographystyle{elsarticle-harv}
\bibliography{references.bib}







\end{document}